\documentclass[sigconf, screen]{acmart}

\usepackage{graphicx}
\usepackage{hyperref}
\AtBeginDocument{%
  }

\copyrightyear{2025}
\acmYear{2025}
\setcopyright{rightsretained}
\acmConference[CHIWORK '25 Adjunct]{CHIWORK '25 Adjunct: Adjunct Proceedings of the 4th Annual Symposium on Human-Computer Interaction for Work}{June 23--25, 2025}{Amsterdam, Netherlands}
\acmBooktitle{CHIWORK '25 Adjunct: Adjunct Proceedings of the 4th Annual Symposium on Human-Computer Interaction for Work (CHIWORK '25 Adjunct), June 23--25, 2025, Amsterdam, Netherlands}
\acmPrice{}
\acmDOI{10.1145/3707640.3731913}
\acmISBN{979-8-4007-1397-2/25/06}

\newcommand{\sysname}{\textsc{ScholarMate}}

\begin{document}

\begin{teaserfigure}
  \centering
  \includegraphics[width=\textwidth]{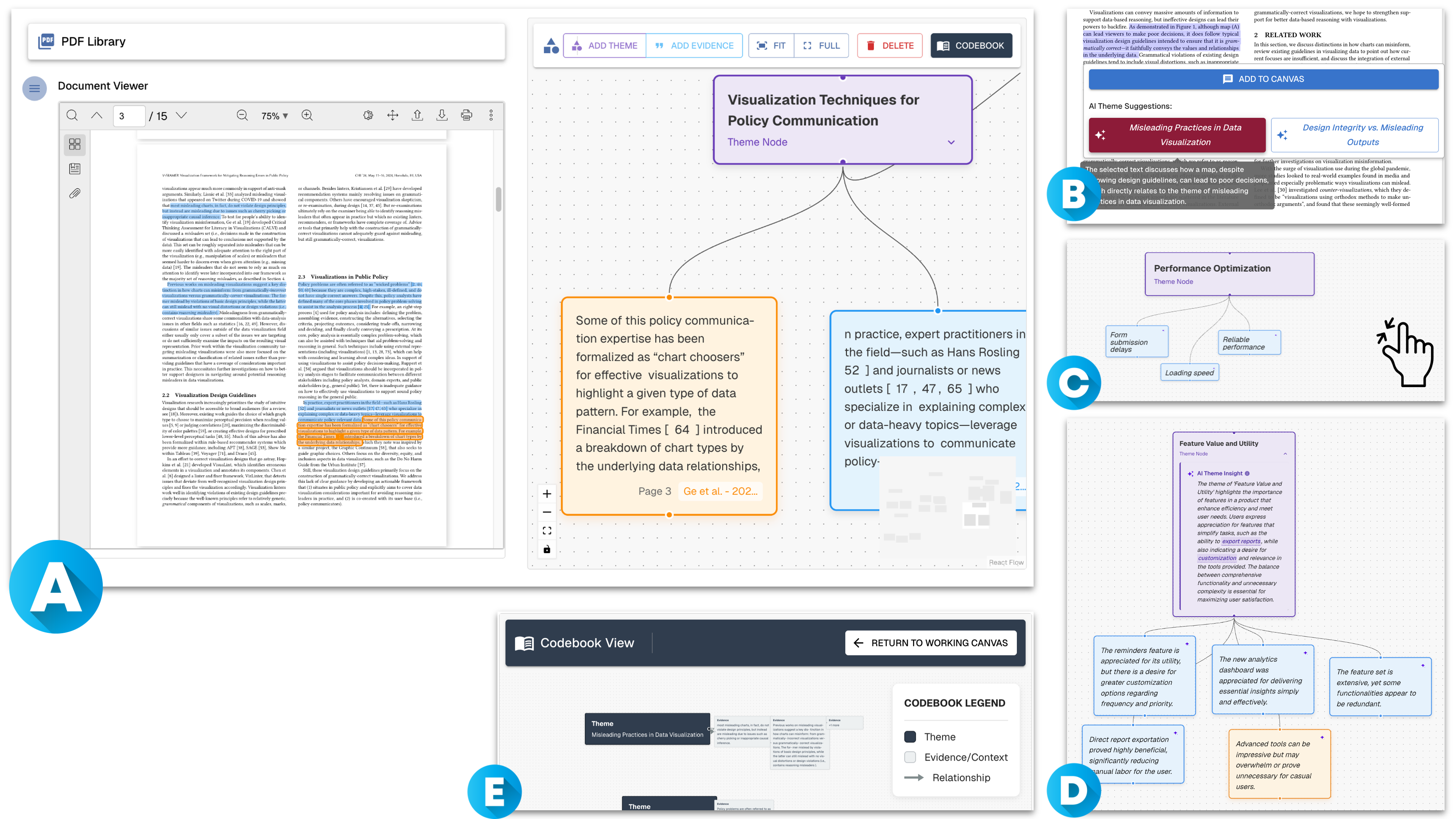}
  \caption{Interface and Key Features of \sysname{}. (A) Researchers view source PDFs (left) alongside an interactive free-form canvas (right), snippets extracted from the source document are visually organized on the canvas while preserving source links;  (B) Mixed-initiative workflow with AI suggestions helps structure textual snippets into themes; (C) Semantic zoom dynamically adjusts the level of detail shown within evidence nodes (from full text to summaries) as the researcher zooms; (D) Theme insight help researcher stay informed during theme development, hoverable insight keyword highlights corresponding evidence(s); (E) Dedicated codebook view summarizes the thematic coding progress. \sysname{} balances direct manipulation with meaningful AI assistance.
  }
  \Description{Interface and Key Features of \sysname{}. (A) Researchers view source PDFs (left) alongside an interactive free-form canvas (right), snippets extracted from the source document are visually organized on the canvas while preserving source links;  (B) Mixed-initiative workflow with AI suggestions helps structure textual snippets into themes; (C) Semantic zoom dynamically adjusts the level of detail shown within evidence nodes (from full text to summaries) as the researcher zooms; (D) Theme insight help researcher stay informed during theme development, hoverable insight keyword highlights corresponding evidence(s); (E) Dedicated codebook view summarizes the thematic coding progress. \sysname{} balances direct manipulation with meaningful AI assistance.}
  \label{fig:teaser}
\end{teaserfigure}

\title{\sysname{}: A Mixed-Initiative Tool for Qualitative Knowledge Work and Information Sensemaking}

\author{Runlong Ye}
\orcid{0000-0003-1064-2333}
\email{harryye@cs.toronto.edu}
\affiliation{
  \institution{Computer Science, \\University of Toronto}
  \city{Toronto}
  \state{Ontario}
  \country{Canada}
}

\author{Patrick Yung Kang Lee}
\orcid{0000-0002-3385-5756}
\email{patricklee@cs.toronto.edu}
\affiliation{
  \institution{Computer Science, \\University of Toronto}
  \city{Toronto}
  \state{Ontario}
  \country{Canada}
}

\author{Matthew Varona}
\orcid{0009-0005-6201-973X}
\email{varona@cs.toronto.edu}
\affiliation{
  \institution{Computer Science, \\University of Toronto}
  \city{Toronto}
  \state{Ontario}
  \country{Canada}
}

\author{Oliver Huang}
\orcid{0009-0007-1585-1229}
\email{oliver@cs.toronto.edu}
\affiliation{
  \institution{Computer Science, \\University of Toronto}
  \city{Toronto}
  \state{Ontario}
  \country{Canada}
}

\author{Carolina Nobre}
\orcid{0000-0002-2892-0509}
\email{cnobre@cs.toronto.edu}
\affiliation{
  \institution{Computer Science, \\University of Toronto}
  \city{Toronto}
  \state{Ontario}
  \country{Canada}
}


\begin{abstract}
Synthesizing knowledge from large document collections is a critical yet increasingly complex aspect of qualitative research and knowledge work. While AI offers automation potential, effectively integrating it into human-centric sensemaking workflows remains challenging. We present \sysname{}, an interactive system designed to augment qualitative analysis by unifying AI assistance with human oversight. \sysname{} enables researchers to dynamically arrange and interact with text snippets on a non-linear canvas, leveraging AI for theme suggestions, multi-level summarization, and evidence-based theme naming, while ensuring transparency through traceability to source documents. Initial pilot studies indicated that users value this mixed-initiative approach, finding the balance between AI suggestions and direct manipulation crucial for maintaining interpretability and trust. We further demonstrate the system's capability through a case study analyzing 24 papers. By balancing automation with human control, \sysname{} enhances efficiency and supports interpretability, offering a valuable approach for productive human-AI collaboration in demanding sensemaking tasks common in knowledge work.
\end{abstract}




\maketitle

\section{Introduction}
Researchers deal with large amounts of text‐based data in their daily workflows, such as reviewing academic publications and conducting qualitative data analysis on interview transcripts. Making sense of these large corpora presents escalating challenges due to the sheer volume of text data being collected and produced. Traditional approaches, such as manual open coding and thematic labeling, require extensive human effort and time, making rapid adjustments to emerging patterns that are challenging when faced with large corpora \cite{braun2006using}. This demanding manual process relies heavily on the researcher's interpretive skill to discern patterns and relationships \cite{LiquidText}, limiting the scale and efficiency of analysis possible \cite{whatisqual}.

Recent advances in large language models (LLMs) have demonstrated their capacity to summarize, cluster, and infer patterns from large textual datasets~\cite{brown2020languagemodelsfewshotlearners}. Despite these capabilities, purely automated solutions can struggle with context awareness and interpretive flexibility, qualities that are critical in qualitative research domains~\cite{10.1145/3703155, hung-etal-2023-walking}. Moreover, integrating AI raises concerns about automation bias, the potential for diminished researcher cognitive engagement \cite{SKITKA1999991}, and the risk of over-reliance on outputs that may lack nuance or contain errors \cite{hou2023should, 10.1145/3703155}. Three core challenges persist in large‐scale AI-assisted analysis: (1) Maintaining transparency and traceability when automated summarization systems are introduced \cite{kryscinski-etal-2020-evaluating}, (2) ensuring researchers remain actively engaged \cite{hou2023should} rather than passively accepting outputs,  and (3) supporting creative insight through flexible exploration of emergent connections during the research process \cite{anderson2024homogenization}.

These challenges lead us to our primary research question: \textbf{\textit{How can an interactive system for qualitative knowledge work effectively weave AI-generated insights with human exploration and critical control, supporting purposeful sensemaking?}} To address this, we follow the human‐centered approach to AI proposed by \citet{shneiderman2020human}, where machines provide computational power while humans guide, interpret, and refine outputs.

We implement this in \textbf{\sysname{}}, an LLM-assisted system designed to support researchers conducting qualitative research work and facilitating the text-based sensemaking process. \sysname{} allows researchers to import a text corpus and manipulate snippets on a non-linear canvas interface, enabling them to build working theories of related documents. This visual re-arrangement process specifically supports human-AI collaborative sensemaking. Furthermore, \sysname{} explores how human-computer interaction (HCI) can shape future work practices involving AI partners, aiming for meaningful and productive outcomes in complex knowledge work.

\section{Related Work}
Our work builds upon prior research exploring computational assistance for qualitative research and text sensemaking. We situate \sysname{} relative to approaches focusing on enhanced manual interaction, AI assistance for specific analytic tasks, and emerging mixed-initiative sensemaking systems.

\subsection{Enhanced Manual Interaction and Visualization}

Prior research on manual interaction and visualization emphasizes fluid user-driven interactions but typically lacks AI capabilities for handling large-scale qualitative analysis. For instance, LiquidText \cite{LiquidText}\footnote{\href{https://www.liquidtext.net/}{https://www.liquidtext.net/}} enables visually connecting document excerpts for active reading and sensemaking. Similarly, visual analytics research has demonstrated meaningful semantic relationships within user-generated spatial layouts \cite{endert2012semantics}. In practice, qualitative researchers frequently use commercial tools such as Miro\footnote{\href{https://miro.com/}{https://miro.com/}}, facilitating manual affinity diagramming through sticky notes to organize textual data into themes \cite{zhang2025blame}. While valuable, these manual systems are limited in efficiency and scalability. \sysname{} directly addresses these gaps by integrating meaningful AI suggestions alongside intuitive manual interaction.

\subsection{AI Assistance for Specific Analytic Tasks} 

Recent systems leveraging AI, particularly large language models (LLMs), often target specific analytic subtasks. IdeaSynth \cite{IdeaSynth}, PersonaFlow \cite{PersonaFlow}, and Sensecape \cite{Sensecape} exemplify AI's role in idea generation, iterative exploration, and multilevel sensemaking. Similarly, DiscipLink \cite{disciplink} and Synergi \cite{synergi} facilitate structured scholarly synthesis but impose constraints on flexible, iterative exploration. These solutions typically lack extensive direct manipulation of textual data, crucial for qualitative analysis. Moreover, \citet{anderson2024homogenization} highlights LLMs' homogenizing effects on ideation, emphasizing the need for diverse interpretations in analytic processes.

\sysname{} significantly extends this prior work by combining AI-driven thematic suggestions and direct manipulation in a unified, flexible interface, uniquely supporting deeper qualitative exploration without constraining analytic processes.

\subsection{Towards Mixed-Initiative Qualitative Sensemaking} 

\sysname{} builds upon research advocating mixed-initiative systems that synergize human expertise and AI. Scholastic \cite{scholastic} uses structured graph-based interactions for collaborative analysis but offers limited spatial and intuitive flexibility. Similarly, \citet{10.1145/3663384.3663398} explore collaborative AI-supported ideation within virtual canvases but primarily in synchronous group contexts, presenting distinct challenges compared to individual or asynchronous qualitative analysis.

\sysname{} innovates by embedding AI within a flexible, canvas-based interface, enhancing interpretability with semantic zoom and dynamically adjusting detail granularity. It explicitly addresses transparency and traceability, linking AI suggestions directly to source documents -- areas identified as critical yet underexplored \cite{amershi2019guidelines, ye2025design}. While other recent systems like LLooM also leverage LLMs for mixed-initiative sensemaking by generating explicit `concepts' \cite{LLooM}, \sysname{} uniquely
bridges direct spatial manipulation valued in manual interaction tools \cite{endert2012semantics, LiquidText} and sophisticated AI assistance essential for large-scale qualitative analysis.

\section{Design Goals}
\label{subsection:design_goals}

Building on our prior design guideline in AI-assisted research \cite{ye2025design}, we outline three primary design goals for 
\sysname{}. These goals address central challenges by carefully balancing AI assistance with human judgment, providing transparent reasoning for AI-generated suggestions, and adhering to ethical standards to ensure trustworthiness. Below, we summarize each design goal's rationale briefly.

\begin{itemize}
    \item \textbf{DG1: Support Mixed-Initiative Human-AI Collaboration.}
While AI can speed up qualitative analysis, over-automation risks undermining critical thinking and nuanced interpretation, potentially causing passive acceptance of flawed outcomes \cite{lee2025impact, passi2022overreliance}, thus, \sysname{} is designed as a mixed-initiative system \cite{10.1145/302979.303030}, aiming to incorporate AI assistance with human expertise. It enables both the user and AI to propose insights while preserving ultimate human control. This design ensures efficiency without compromising essential human judgment and agency~\cite{scholastic} in the knowledge co-creation process.
\item \textbf{DG2: Ensure Interpretability of AI Reasoning.}
AI models often act as opaque "black boxes," which can hinder researchers' ability to critically assess their outputs and reasoning processes \cite{miller2018explanationartificialintelligenceinsights, ye2025design}. To illuminate the AI's reasoning process, \sysname{} provides transparency into the basis of each suggestion. It explicitly links AI-generated suggestions to the specific evidence snippets that prompted them, offering a direct rationale. A visual preview further clarifies the immediate impact of accepting a suggestion. Making this rationale transparent allows researchers to critically evaluate the logic behind a suggestion, facilitating informed interaction: acceptance, revision, or rejection, based on their own understanding \cite{10.1145/2678025.2701399}. This aligns with human-AI interaction guidelines promoting system intelligibility \cite{amershi2019guidelines}.
\item \textbf{DG3: Promote Output Validation and Ethical Use.}
AI tools can introduce errors through biases or hallucinations, posing ethical challenges and threatening research validity \cite{ye2025design, holstein2019improving}. Therefore, \sysname{} incorporates features focused on validating AI-generated content and ensuring ethical integration. Direct traceability allows researchers to easily verify AI claims or summaries against original source documents, confirming accuracy. Transparently communicating AI limitations and mandating human oversight are crucial mechanisms to mitigate risks associated with potential errors or biases. This emphasis on verification and human control helps maintain ethical standards and fosters justified trust in the final research outcomes.
\end{itemize}

\section{Probe Design}

\sysname{}'s design was informed by observing qualitative researchers' existing practices. We analyzed how they used informal spatial arrangements and annotations (often sketches) to cluster document excerpts and build understanding during their real-world workflows. Insights from these observations, combined with iterative internal design reviews focused on addressing identified researcher needs, guided the development of successive low- and high-fidelity prototypes. 

This process led to the current web-based technology probe \cite{10.1145/642611.642616}, to demonstrate the initial functionalities of \sysname{}. The \sysname{} probe can be \href{https://scholar-mate-eta.vercel.app/}{accessed online}\footnote{\href{https://scholar-mate-eta.vercel.app/}{https://scholar-mate-eta.vercel.app/}}. \sysname{} is built using NextJS, with \texttt{react-pdf-viewer} and \texttt{ReactFlow} components, and designed to support interactive, mixed-initiative qualitative analysis through a flexible, non-linear canvas-based interface. It addresses challenges researchers face in systematically analyzing data while balancing automated assistance with human analytic control. Grounded in mixed-initiative interaction principles \cite{10.1145/302979.303030} and established qualitative research methodologies \cite{braun2006using}, \sysname{} ensures automated suggestions remain transparent and subject to human validation, recognizing the critical role of human judgment in resolving ambiguities and refining interpretations.

Researchers begin by uploading their corpus into \sysname{}, enabling them to extract relevant text snippets via a built-in document viewer directly onto an interactive canvas. The choice of this canvas-based model aligns with literature findings on spatial document clusterization \cite{endert2012semantics}, indicating spatial organization of textual data can meaningfully support researchers in discerning thematic structures in their corpus. This approach proves particularly valuable during exploratory analysis, where ambiguity about themes is most prevalent.

\subsection{Interactive Canvas}




Within this canvas, extracted snippets become \textit{manipulable nodes}. \sysname{} distinguishes between \textit{evidence nodes}, containing direct excerpts or annotations, and \textit{theme nodes}, used to visually cluster related evidence. Researchers maintain full agency to create, rearrange, connect, edit, and delete nodes, facilitating an iterative process of structuring and refining their analysis. This direct manipulation ensures researchers retain control over data organization and interpretation.

To navigate potentially complex canvas layouts while preserving relative spatial positioning of data, \sysname{} incorporates \textit{Semantic Zoom}. Grounded in visualization principles like Shneiderman's mantra -- \textit{``overview first, zoom and filter, then details on demand''}~\cite{shneiderman2003eyes} -- for exploring large datasets, this allows fluid transitions between high-level thematic overviews and detailed evidence views. As a researcher zooms out, evidence text transforms into progressively shorter, LLM-generated summaries of original content. Conversely, zooming in reveals longer summaries or original text excerpts. While evidence nodes adapt their content detail, theme nodes primarily scale visually with the zoom level, retaining their fixed text to ensure stable thematic anchors.

This combination of direct manipulation and AI-driven semantic zoom primarily supports \textit{Mixed-Initiative Collaboration (DG1)} by blending user control with AI assistance. Furthermore, the crucial ability to trace any evidence snippet on the canvas directly back to its context in the original source document promotes \textit{Output Validation (DG3)}, allowing researchers to easily verify information and trust the evidence underpinning their analysis.

\subsection{Working Canvas and Codebook View}
\sysname{} provides two distinct but interconnected interfaces: the Working Canvas and the Codebook View. Both interfaces utilize \texttt{ReactFlow} component, semantic zoom, node-based interactions, and maintain direct evidence-to-PDF highlighting capabilities. 

The \textit{Working Canvas} is designed to support active data analysis, enabling users to freely organize, edit, and refine evidence and theme nodes. It supports full interactivity, allowing detailed modifications, annotations, and iterative analysis, empowering users to co-create with AI, and dynamically explore emerging thematic insights. Conversely, the \textit{Codebook View} offers a stable, read-only visualization of current themes and evidence. To facilitate comprehension without overwhelming users, the Codebook View limits the display to two pieces of evidence per theme. This interface emphasizes high-level theme exploration, providing a quick reference for users to check their understanding or synthesize existing themes.

This dual-interface design directly enhances \textit{Mixed-Initiative Collaboration (DG1)} by providing distinct spaces for dynamic, AI-supported exploration versus human-led refinement and finalization. Separating exploratory work (Working Canvas) from the stable, read-only synthesis (Codebook View) also promotes \textit{Ethical Use (DG3)} by improving the accountability and trustworthiness of the final thematic structure.

\subsection{AI-Assisted Features}
\sysname{} integrates AI seamlessly throughout the analytical process, enhancing researchers' efficiency and depth of insight. Key AI-assisted functionalities include:

\textit{Theme Suggestions:} After a user selects a text excerpt, \sysname{} analyzes it and the existing workspace themes. It then suggests either assigning the excerpt to a relevant existing theme or using it to establish a new one. Alternatively, users can choose to manually incorporate the selected excerpt directly onto the workspace, allowing them full control over its placement and connections within their analysis structure. AI suggestions are visibly distinguished by special UI indicators and transparently explained, supporting informed choices and ensuring researchers maintain overall control over their thematic decisions, whether leveraging AI assistance or proceeding manually.

\textit{Multi-level Semantic Summarization:} \sysname{} employs multi-tiered summarization (medium, short, and tiny) on top of the original passage from context nodes and dynamically adapts to zoom levels on the canvas and functionally supports the semantic zoom feature. This ensures optimal readability and contextual relevance at varying scales.

\textit{Contextual Theme Naming:} Researchers can utilize AI-generated theme naming functionality that suggests theme names based on the aggregated context of connected evidence nodes. This feature supports the iterative nature of qualitative research by allowing researchers to continually refine theme names as their understanding of the data evolves. It helps researchers quickly capture and communicate the essence of thematic clusters, enhancing conceptual coherence and facilitating iterative sense-making.

\textit{Evidence-Grounded Theme Descriptions:} Leveraging AI, \sysname{} generates theme descriptions that automatically incorporate salient, evidence-derived keywords. These keywords are interactively linked to the source evidence nodes. Hover interactions provide immediate visual feedback by highlighting these connections, directly grounding the abstract theme description in concrete data and enhancing clarity through traceable AI insights.

These AI-assisted features embody \textit{Mixed-Initiative Collaboration (DG1)} by offering suggestions while preserving user autonomy. Features like Theme Suggestions and Evidence-Grounded Theme Descriptions directly address \textit{Interpretability of AI Reasoning (DG2)} by linking suggestions and keywords to the underlying evidence, explaining why they were proposed. Concurrently, the reliance on source evidence for suggestions and summaries necessitates and supports \textit{Output Validation (DG3)}, encouraging researchers to verify AI contributions against original documents.

\section{Evaluation}
We evaluate \sysname{} through a small pilot study as well as a use case with a real-world corpus of papers.
\subsection{Pilot Study}
We conducted a preliminary evaluation of \sysname{} through a small pilot study with two graduate students (P1, P2), both with backgrounds in human-computer interaction and data visualization. We mainly aim to understand: (1) participants' strategies for using \sysname{} in a qualitative research task and (2) their perceptions of its usefulness. Participants were asked to complete a thematic analysis task on two familiar research papers \cite{DracoGPT, VizAbility}, pre-loaded into \sysname{} with researcher-provided highlights matching two predefined themes. Participants, however, were not informed of these pre-identified themes and were asked to independently identify themes using \sysname{}. Afterward, we administered NASA-TLX \cite{NASATLX} to measure perceived cognitive load and UMUX-LITE \cite{10.1145/2470654.2481287} to assess 
system usability, followed by brief semi-structured interviews for qualitative feedback.

\subsection{Quantitative Results} While our sample size was small, participants rated \sysname{} highly, with an average satisfaction score of 4.5 out of 5. The NASA-TLX scores indicated low-moderate cognitive demands (aggregated load: 2.41/7), primarily driven by low frustration (average 1.5/7), physical demand (average 1/7), and high perceived task performance (average 6/7). UMUX-LITE results showed strong usability ratings, indicating high participant agreement that the system met their requirements and was easy to use.

\subsection{Qualitative Results} 
Participants identified both strengths and areas for improvement:

\subsubsection{Beneficial AI Integration}
Participants found the AI suggestions for themes and insights helpful, particularly for managing the coding process. P2 noted the utility of AI suggestions, especially when initial ideas weren't clear: \textit{``I used the AI suggestion when it felt like none of the existing ones fit particularly well'' (P2)}. The AI's ability to suggest existing themes was seen as a significant time-saver: \textit{``If this thing can consistently assign the right [theme] that saves me a ton of time'' (P2)}. New AI-generated themes were considered valuable starting points, provided they could be modified: \textit{``I think it's still useful to have that option [to use an AI-suggested theme and edit it later]'' (P2)}. P1 also acknowledged the AI's capability at the initial stage: \textit{``For each of the highlighted texts... they can do separate things very well'' (P1)}.

\subsubsection{Freeform Interaction}
The canvas-based, flexible nature of organizing themes was appreciated. P2 liked the different ways to categorize highlights: \textit{``There's like all of the modes of categorizing stuff because you have the you can just add to the canvas and leave the categorization free later or make a manual theme node'' (P2)}. This flexibility contributed to a positive overall impression: \textit{``it seems like a way I would actually want to do qualitative research'' (P2)}. Participants also took vastly different approaches to organizing their evidence and theme nodes: P1 generated a multi-layered tree structure with sub-themes, while P2 preferred a flat structure where multiple themes were combined into larger ones.

\subsubsection{Limitation and Future Improvments}
\begin{itemize}
    \item \textbf{Usability:} P2 desired better node management: \textit{"[It would be useful] maybe being able to like drag multiple nodes at once... it's useful to not have to drag every single one" (P2)}. Scaling was also a concern for real-world use: \textit{``the scale, I think, is super different for an actual project... whether it's easy to manage like multiple nodes'' (P2)}.
    \item \textbf{AI Limitations:} P1 observed that the insight generation might not properly account for sub-theme structures: \textit{``if I click 'generate insight', they still [don't seem to prioritize the hierarchy]... they don't take too much attention to the sub [themes], like the hierarchical [themes] underneath''} (P1).
    \item \textbf{Feature Requests:} P1 wished for an undo or state history feature: \textit{``sometimes I want to perhaps go back to my last stage... I don't know how to go back to the original one after I combined it'' (P1)}. P1 suggested a chat interface: \textit{``perhaps there could be a like a little chat bot... So sometimes I want to combine certain things together. I can ask [the] AI for suggestions before I actually combine those'' (P1)}. Both participants discussed the potential for a "top-down" approach, allowing users to guide the AI: \textit{``maybe before I even reading this paper... I can input a general requirement I'm looking for in those papers'' (P1)}, or \textit{``give a prompt saying, Hey, this is generally what I'm interested in'' (P2)}.
\end{itemize}

\subsection{Use Case: Analyzing Vector Database Papers with \sysname{}}

We also evaluate \sysname{}'s capabilities through a use case analyzing 24 state-of-the-art vector database papers. We reference the thematic structure established in a graduate seminar course.  We utilized the system to automatically generate themes and classify papers using abstracts. \sysname{} showed moderate alignment (7/16 correctly classified) after the first iteration. Interactively refining the system-generated themes substantially improved categorization accuracy (13/16 correctly classified), with higher accuracy achievable by incorporating additional excerpts.

Notably, \sysname{} is not limited to matching predefined categories. Instead, its primary value is enabling iterative sensemaking and thematic exploration. The interactive workspace and semantic zoom capabilities empowered researchers with fluid transitions, moving smoothly from macro-level thematic overviews to micro-level analysis of individual excerpts, highlighting \sysname{}'s strength in supporting sophisticated analysis.

\section{Limitations and Future Work}
As \sysname{} incorporates interactions with LLMs, it faces inherent risks of AI error. To reduce over-reliance on AI, we designed \sysname{} to foster cognitive engagement by providing direct links to source excerpts, explanations for suggested themes, and easy-to-use editing capabilities. Future iterations could enhance the analytical workflow by supporting more complex, realistic hierarchical structures (e.g., themes, sub-themes, nested concepts) to better capture nuanced relationships within the data. Building upon existing interpretability features, we plan to investigate various interaction designs for human-AI collaboration, evaluating how they influence appropriate trust and reliance based on recent insights into user interactions with LLMs \cite{bo2025}. Additionally, while initial feedback is promising, larger-scale user studies involving diverse researchers and analysts are essential to rigorously evaluate \sysname{}'s interactions, scalability, and impacts on real-world practices. Finally, to further streamline analysis, future work could integrate more automated tagging features guided by user-defined codebooks and implement enhanced privacy controls for sensitive data.

\section{Conclusion}
We introduced \sysname{}, an LLM-assisted visual interactive tool enhancing qualitative analysis through a mixed initiative canvas. Guided by design goals of supporting mixed-initiative human-AI collaboration (DG1), ensuring interpretability of AI reasoning (DG2), and promoting output validation and ethical use (DG3), \sysname{} aims to foster effective human-AI partnerships that augment, rather than automate, the nuanced process of qualitative research.  Our initial case study and pilot evaluations suggest that \sysname{} can serve as a useful tool for textual data exploration and organization, successfully supporting sensemaking while maintaining the researcher's critical cognitive engagement.

Beyond its primary function in thematic analysis, the flexible architecture of \sysname{} by combining spatial organization, semantic zoom, node-based interactions, and integrated AI assistance, points towards compelling possibilities for broader scholarly applications. We envision researchers leveraging its interactive canvas for tasks such as \textit{visualizing the conceptual structure of literature reviews}, spatially mapping relationships between key publications or concepts, identifying thematic connections across disparate sources, and pinpointing research gaps more systematically. Similarly, \sysname{} could be adapted for \textit{analyzing argumentation structures}, allowing users to map claims, counterclaims, and supporting evidence extracted from scholarly texts. Furthermore, it holds potential as a \textit{dynamic scholarly notebook or personal knowledge management tool}, enabling researchers to visually organize and interconnect notes, emergent ideas, and textual evidence across projects, thereby allowing for conceptual synthesis and supporting the potential for emergent insights.

\sysname{} offers insights for designing future human-AI collaborative work environments, where technology not only enhances productivity and expedites analytical tasks, but also deepens researchers' analytical engagement with their data and ideas. Ultimately, we aim to facilitate purposeful, transparent, fluid, and trustworthy human-AI partnerships in the pursuit of knowledge.

\begin{acks}
We would like to thank Prof. \href{https://orcid.org/0000-0003-2965-5302}{Michael Liut} for supporting this work.

\noindent We acknowledge the support of the Natural Sciences and Engineering Research Council of Canada (NSERC), [funding reference number RGPIN-2024-04348 and RGPIN-2024-06005].
\end{acks}

\bibliographystyle{ACM-Reference-Format}
\balance
\bibliography{reference}



\end{document}